\documentclass[a4paper,twocolumn,10pt,twoside]{IEEEtran}
\usepackage[utf8]{inputenc}

\usepackage{graphicx,epic,eepic,epsfig,amsmath,latexsym,amssymb,verbatim,xcolor}

\usepackage{amstext}
\usepackage{dsfont}
\usepackage{cite}

\usepackage[colorlinks=true, linkcolor=blue, urlcolor=blue,citecolor=blue]{hyperref}

\usepackage{orcidlink}

\usepackage{theorem}
\newtheorem{definition}{Definition}

\newtheorem{lemma}[definition]{Lemma}

\newtheorem{theorem}[definition]{Theorem}
\newtheorem{corollary}[definition]{Corollary}

\def\squareforqed{\hbox{\rlap{$\sqcap$}$\sqcup$}}
\def\qed{\ifmmode\squareforqed\else{\unskip\nobreak\hfil
\penalty50\hskip1em\null\nobreak\hfil\squareforqed
\parfillskip=0pt\finalhyphendemerits=0\endgraf}\fi}
\def\endenv{\ifmmode\;\else{\unskip\nobreak\hfil
\penalty50\hskip1em\null\nobreak\hfil\;
\parfillskip=0pt\finalhyphendemerits=0\endgraf}\fi}

\mathchardef\ordinarycolon\mathcode`\:
\mathcode`\:=\string"8000
\def\vcentcolon{\mathrel{\mathop\ordinarycolon}}
\begingroup \catcode`\:=\active
  \lowercase{\endgroup
  \let :\vcentcolon
  }

\newcommand{\nc}{\newcommand}
\nc{\rnc}{\renewcommand}
\nc{\beq}{\begin{equation}}
\nc{\eeq}{{\end{equation}}}
\nc{\beqa}{\begin{eqnarray}}
\nc{\eeqa}{\end{eqnarray}}
\nc{\lbar}[1]{\overline{#1}}
\nc{\bra}[1]{\langle#1|}
\nc{\ket}[1]{|#1\rangle}
\nc{\ketbra}[2]{|#1\rangle\!\langle#2|}
\nc{\braket}[2]{\langle#1|#2\rangle}
\nc{\proj}[1]{| #1\rangle\!\langle #1 |}
\nc{\avg}[1]{\langle#1\rangle}
\nc{\Rank}{\operatorname{Rank}}
\nc{\smfrac}[2]{\mbox{$\frac{#1}{#2}$}}
\nc{\tr}{\operatorname{Tr}}
\nc{\ox}{\otimes}
\nc{\dg}{\dagger}
\nc{\dn}{\downarrow}
\nc{\cA}{{\cal A}}
\nc{\cB}{{\cal B}}
\nc{\cC}{{\cal C}}
\nc{\cD}{{\cal D}}
\nc{\cE}{{\cal E}}
\nc{\cF}{{\cal F}}
\nc{\cG}{{\cal G}}
\nc{\cH}{{\cal H}}
\nc{\cI}{{\cal I}}
\nc{\cJ}{{\cal J}}
\nc{\cK}{{\cal K}}
\nc{\cL}{{\cal L}}
\nc{\cM}{{\cal M}}
\nc{\cN}{{\cal N}}
\nc{\cO}{{\cal O}}
\nc{\cP}{{\cal P}}
\nc{\cQ}{{\cal Q}}
\nc{\cR}{{\cal R}}
\nc{\cS}{{\cal S}}
\nc{\cT}{{\cal T}}
\nc{\cU}{{\cal U}}
\nc{\cW}{{\cal W}}
\nc{\cX}{{\cal X}}
\nc{\cY}{{\cal Y}}
\nc{\cZ}{{\cal Z}}
\nc{\csupp}{{\operatorname{csupp}}}
\nc{\qsupp}{{\operatorname{qsupp}}}
\nc{\var}{{\operatorname{var}}}
\nc{\Var}{{\operatorname{Var}}}
\nc{\rar}{\rightarrow}
\nc{\lrar}{\longrightarrow}
\nc{\polylog}{{\operatorname{polylog}}}
\nc{\1}{\mathds{1}}
\nc{\wt}{{\operatorname{wt}}}
\nc{\av}[1]{{\left\langle {#1} \right\rangle}}

\nc{\RR}{{{\mathbb R}}}
\nc{\CC}{{{\mathbb C}}}
\nc{\FF}{{{\mathbb F}}}
\nc{\NN}{{{\mathbb N}}}
\nc{\ZZ}{{{\mathbb Z}}}
\nc{\PP}{{{\mathbb P}}}
\nc{\QQ}{{{\mathbb Q}}}
\nc{\UU}{{{\mathbb U}}}
\nc{\EE}{{{\mathbb E}}}
\nc{\id}{{\operatorname{id}}}

\nc{\CHSH}{{\operatorname{CHSH}}}

\nc{\be}{\begin{equation}}
\nc{\ee}{{\end{equation}}}
\nc{\bea}{\begin{eqnarray}}
\nc{\eea}{\end{eqnarray}}
\nc{\<}{\langle}
\rnc{\>}{\rangle}
\nc{\Hom}[2]{\mbox{Hom}(\CC^{#1},\CC^{#2})}
\nc{\rU}{\mbox{U}}

\nc{\ob}[1]{#1}

\nc{\SEP}{{\text{SEP}}}
\nc{\NS}{{\text{NS}}}
\nc{\LOCC}{{\text{LOCC}}}
\nc{\PPT}{{\text{PPT}}}
\nc{\EXT}{{\text{EXT}}}
\nc{\Sym}{{\operatorname{Sym}}}

\nc{\ERLO}{{E_{\text{r,LO}}}}
\nc{\ERLOCC}{{E_{\text{r,LOCC}}}}
\nc{\ERPPT}{{E_{\text{r,PPT}}}}
\nc{\ERLOCCinfty}{{E^{\infty}_{\text{r,LOCC}}}}
\nc{\Aram}{{\operatorname{\sf A}}}

\begin{document}

\title{Entropic proofs of Singleton bounds for quantum error-correcting codes}

\author{Markus Grassl${}^{\orcidlink{0000-0002-3720-5195}}$, \IEEEmembership{Senior Member, IEEE}, 
  Felix Huber${}^{\orcidlink{0000-0002-3856-4018}}$, and 
  Andreas Winter${}^{\orcidlink{0000-0001-6344-4870}}$
\thanks{
Markus Grassl is with the International Centre for Theory of Quantum
Technologies, University of Gdansk, 80-308 Gda\'nsk, Poland.}
\thanks{
Felix Huber was with ICFO -- Institut de Ci\`encies Fot\`oniques,
ES-08860 Castelldefels (Barcelona), Spain, and is now with
the Atomic Optics Department, Jagiellonian University, 30-348 Krak\'ow, Poland.}
\thanks{
Andreas Winter is with the Departament de F\'isica: Grup
d'In\-for\-maci\'o Qu\`antica, Universitat Aut\`onoma de Barcelona,
ES-08193 Bellaterra (Barcelona), Spain, as well as with ICREA --
Instituci\'o Catalana de Recerca i Estudis Avan\c{c}ats, Pg. Lluis
Companys, 23, 08010 Barcelona, Spain.}
\thanks{
This paper was presented at Beyond IID in Information Theory 8, Stanford, 9--13 November 2020.
}
\bigskip\bigskip
}

\maketitle

\markboth{IEEE Transactions on Information Theory,
  \MakeLowercase{accepted for publication}, 2022}{Grassl \MakeLowercase{\textit{et al.}}: Entropic proofs of Singleton bounds for quantum error-correcting codes}

\begin{abstract}
We show that a relatively simple reasoning using von Neumann entropy
inequalities yields a robust proof of the quantum Singleton bound for
quantum error-correcting codes (QECC).  For entanglement-assisted
quantum error-correcting codes (EAQECC) and catalytic codes (CQECC), a
type of generalized quantum Singleton bound [Brun \textit{et al.}, IEEE
Trans. Inf. Theory 60(6):3073--3089 (2014)] was believed to hold for
many years until recently one of us found a counterexample [MG,
Phys. Rev. A 103, 020601 (2021)]. Here, we rectify this state of
affairs by proving the correct generalized quantum Singleton bound,
extending the above-mentioned proof method for QECC; we also 
prove information-theoretically tight bounds on the entanglement-communication 
tradeoff for EAQECC. 
All of the bounds relate block length $n$ and code length $k$ for given minimum 
distance $d$ and we show that they are robust, in the sense that they hold with 
small perturbations for codes which only correct most of the erasure errors of less than $d$ letters. 
In contrast to the classical case, the bounds take on qualitatively different forms depending 
on whether the minimum distance is smaller or larger than half the block length.
We also provide a propagation rule: any pure QECC yields an
EAQECC with the same distance and dimension, but of shorter block
length.
\end{abstract}


\section{Introduction}
\begin{figure}[ht]
  \centerline{\includegraphics[width=\hsize]{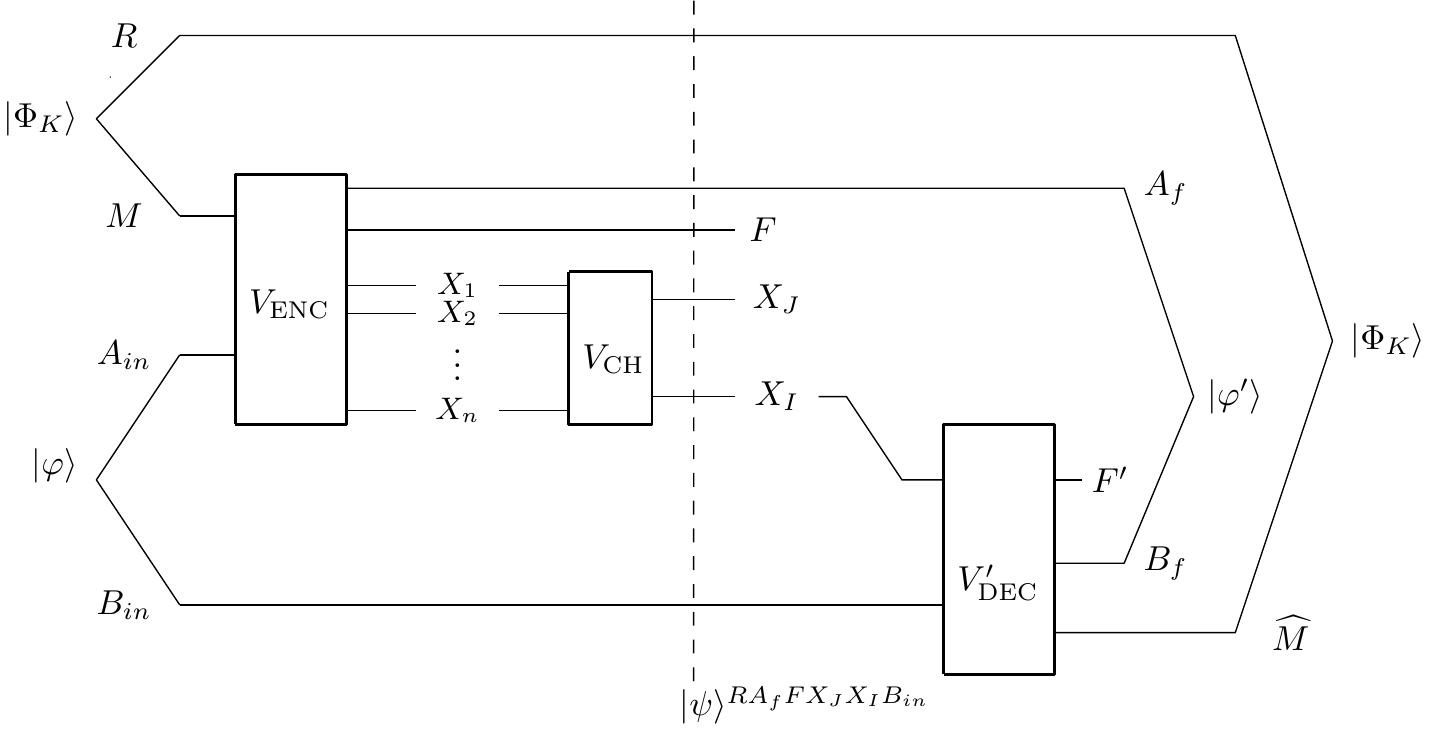}}
  \caption{Diagramme of a general entanglement-assisted quantum
    error-correcting code (EAQECC) as a quantum transmission and
    entanglement-generation procedure between the encoder
    $V_{\text{ENC}}$ and the decoder $V'_{\text{DEC}}$.  Both encoder
    and decoder are written as isometries, with auxiliary output
    systems $F$ and $F'$, respectively. The erasure channel
    $V_{\text{CH}}$ partitions the systems $X_1 X_2\ldots X_n$ into
    two disjoint blocks $X_I$ and $X_J$ with $I,J \subset[n]$ that are
    of size $n-d+1$ and $d-1$ respectively. The block $X_J$ is erased,
    while the block $X_I$, together with information about the index
    set $I$, is received by the decoder. We usually assume that all
    $X_i$ have the same dimension~$q$.}
  \label{fig:EAQECC}
\end{figure}

\IEEEPARstart{T}{he} object of the present paper are quantum
error-correcting codes (QECC), plain and with entanglement assistance.
The general communication diagramme is shown in Fig.~\ref{fig:EAQECC}:
a $K$-dimensional system $M$ (the ``message'') is encoded into $n$
quantum systems $X_1,\ldots,X_n$ having $q$ levels each.  (Note that
we do not restrict $q$ to be power of prime).  This is assisted by
entanglement $\ket{\varphi}^{A_{in}B_{in}}$ at the input, which is
quantified by the entanglement entropy $\ell_{in} = S(A_{in})_\varphi
= S(\varphi^{A_{in}}) = S(B_{in})_\varphi$, where we denote by
$\varphi^S$ the marginal state of $\varphi$ on some system $S$.  A
QECC of minimum distance $d$ can correct up to $d-1$ erasures
\cite{GrasslBethPellizarri1997}.  In this case, the quantum erasure
channel partitions the $n$ input systems $X_1\ldots X_n$ into two
disjoint blocks $X_I$ and $X_J$ of size $|I|=n-d+1$ and $|J|=d-1$,
respectively. The decoder receives only the block $X_I$, together with
classical information on the index set $I\subset \{1,\ldots,n\}=[n]$.
The perfect decoding of the quantum information $M$ on the output
system $\widehat{M}$ is equivalent to reproducing the maximally
entangled state $\ket{\Phi_K}^{MR}$ between $\widehat{M}$ and $R$
(i.\,e.~$\ket{\Phi_K}^{\widehat{M} R}$) with fidelity $1$, where $R$
is a reference system.  Additionally, we allow the recovery of some of
the entanglement between the subsystems $A_f$ and $B_f$ in a pure
state $\ket{\varphi'}^{A_{f}B_{f}}$, which is quantified as $\ell_{f}
= S(A_{f})_{\varphi'}$.  The net consumption $\ell := \ell_{in} -
\ell_{f}$ of entanglement can thus be positive, negative, or zero.
The amount of quantum information encoded into $M$ is $\log_2 K$.
Note that in information theory, $\log$ is usually understood to be
the binary logarithm, in particular in the von Neumann entropy
$S(\rho)=-\tr\rho\log_2\rho$.  This convention means that the quantum
information is counted in units of qubits, and the entanglement in
units of ebits.  However, in coding theory, assuming that all $X_i$
are $q$-dimensional, the $q$-ary logarithm is preferred. To avoid
confusion, we will include the base in the notation for the logarithm
throughout.

One of the basic bounds in coding theory is the Singleton
bound~\cite{singleton}.  It has a well-known quantum analogue due to
Knill and Laflamme~\cite{q-singleton}, upper bounding the dimension
$K$ in the absence of any prior entanglement (see also
\cite{Rains,Klappi}). For an unassisted code, i.\,e.~$\ell_{in}=0$, it states
in our notation that
\begin{equation}
  \log_q K \leq n-2d+2,
\end{equation}
which in particular means that $d-1 < \frac{n}{2}$ is necessary for
any non-trivial code with $K>1$ to exist (indeed, this holds due to the
familiar no-cloning argument \cite{BDSW}).
We start by reviewing a proof of the quantum Singleton bound that uses
a simple entropic reasoning~\cite{HuberGrassl}, and that gives a
potentially tighter entropic bound for unassisted codes than the one
previously known~\cite{q-singleton, Rains}.  The bulk of the paper is
however concerned with entanglement-assisted codes
\cite{BrunDevetakHsieh-EA}, for which we generalize this first proof.

In actual codes it will often be the case that $K$ is a power of $2$
(or of $q$), and also $\ket{\varphi}$, $\ket{\varphi'}, \ket{\Phi}$
will be maximally entangled states of Schmidt rank a power of $2$ (or
of $q$).  In that case, $\log_2 K$ counts the number of qubits
encoded; similarly, $\ell_{in}$ and $\ell_{f}$ count the initial and
final number of EPR pairs (ebits), respectively.  But to achieve the
largest generality of our bounds, which state necessary conditions for
the existence of codes, we shall make no such assumption and none of
$\log_2 K$, $\log_q K$, $\ell_{in}$, and $\ell_{f}$ needs to be an
integer.

\begin{figure}[ht]
  \centerline{\includegraphics[width=\hsize]{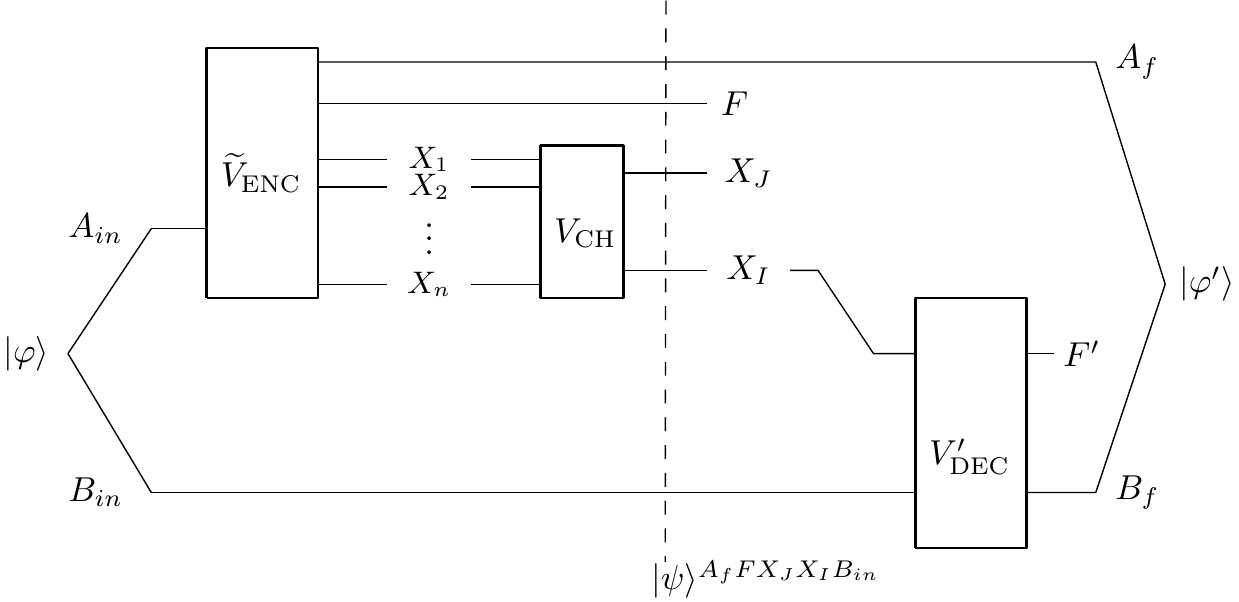}}
  \caption{Diagramme of a general entanglement-assisted quantum error-correcting 
           code (EAQECC) as an entanglement-generation procedure between the
           encoder and the decoder.  Starting from
           Fig.~\ref{fig:EAQECC}, we simply integrate the creation
           of the maximally entangled state $\ket{\Phi_K}^{RM}$ into the
           encoding isometry $\widetilde{V}_{\text{ENC}}$ and merge $R$ and
           $A_{f}$ to get a larger register for the generated entanglement.}
  \label{fig:leaf}
\end{figure}

Since the code can transmit $\log_2 K$ qubits from Alice to Bob, it
can also be used to distribute $\log_2 K$ ebits of entanglement. In
fact, in the communication diagramme, it would be Alice to prepare the
maximally entangled Bell state $\ket{\Phi_K}^{RM}$, keeping $R$ and
encoding $M$. This shifts the focus from sending qubits to generating
entanglement between the parties: starting from an initial state
$\ket{\varphi}^{A_{in}B_{in}}$ and ending with
$\ket{\Phi_K}^{R\widehat{M}} \otimes \ket{\varphi'}^{A_{f}B_{f}}$, the
net entanglement generated is $\log_2 K-\ell$ ebits.  This is the
point of view of catalytic quantum error
correction~\cite{BrunDevetakHsieh-catalytic}.  It allows us to
simplify the communication diagramme: the generation of $\ket{\Phi_K}$
becomes part of the encoding isometry, while renaming $A_{f}R$ as
$A_{f}$ and $B_{f}\widehat{M}$ as $B_{f}$; we thus arrive at
Fig.~\ref{fig:leaf}. This procedure turns any entanglement-assisted
quantum error-correcting codes (EAQECC) for $\log_2 K$ qubits and
using $\ell$ ebits of entanglement into a catalytic quantum
error-correcting code (CQECC) with net generation of $\log_2 K-\ell$
ebits.  Naturally, for such catalytic code to be useful one wants
$\log_2 K-\ell>0$.

\medskip
The rest of the paper is structured as follows: In Section
\ref{sec:q-Singleton}, as a warm-up we re-derive the quantum Singleton
bound for subspace codes by a simple entropic reasoning. In Section
\ref{sec:EAQECC-Singleton} we show that its proof generalizes to a
proof of a corrected version of the originally claimed quantum
Singleton bound for general EAQECCs and CQECCs, while in Section
\ref{sec:full-tradeoff}, we prove tight tradeoff relations between
encoded quantum information and net entanglement consumption in
general EAQECCs.  In Section \ref{sec:constructions} we present
constructions, for sufficiently large alphabet size, showing the
tightness of the derived bounds, and conclude in Section \ref{sec:disc}.

\section{Singleton bound\protect\\ for isometric-encoding QECC}
\label{sec:q-Singleton}
We start with the usual setting of QECC as subspaces of the $n$-party system
$X_1 X_2\ldots X_n$, where the encoder does not share prior entanglement. That is, 
in Figs.~\ref{fig:EAQECC} and~\ref{fig:leaf}, the subsystems $A_{in}$,
$B_{in}$ and $F$ are trivial ($F = A_{in} = B_{in} = \CC$). 

Define by
\begin{alignat}{5}
\overline\sigma = \frac{1}{d-1}\EE_{|J|=d-1} S(X_{J}) \leq \log_2 q \label{eq:average entropy}
\end{alignat}
the average entropy of a random $(d-1)$-block per system.  Note that
in keeping with coding convention, we will from now on focus on
situations where all systems are $q$-ary.

\begin{theorem}[Entropic quantum Singleton bound \cite{HuberGrassl}]
\label{thm:singleton} 
Let $\mathcal{Q}$ be a subspace of $X_1 X_2\ldots X_n$ that
corresponds to a QECC $(\!(n,K,d)\!)_q$ of 
dimension $K>1$ and distance $d$. Then, 
\begin{align}
  \log_2 K 
  &\leq \max\{0,n-2d+2\} \overline\sigma \notag\\ 
  &\leq \max\{0,n-2d+2\} \log_2 q \,.\label{eq:entropic_bound}
\end{align}
\end{theorem}

\begin{IEEEproof}
Purify the maximally mixed state on the code space, 
$\rho = \frac{1}{K}\Pi_Q$ 
where $\Pi_Q$ is the projector onto the code space, 
with a reference system~$R$ of dimension $K$, and go to the entanglement-generating 
code picture (thus, $A_{in}=B_{in}=\CC$ in Fig.~\ref{fig:EAQECC}, or equivalently
$R \equiv A_{f}$ in Fig.~\ref{fig:leaf}, while $A_{in}=B_{in}=\CC$).
The condition for perfect quantum error correction 
is equivalent, for a code of distance $d$, to the relation
\begin{equation}\label{eq:KLF_in_entropies}
  S(X_I) = S(R X_J) = S(R) + S(X_J)
\end{equation} 
for any bipartition $[n] = I \stackrel{.}{\cup} J$ into disjoint subsets of
cardinalities $|I| = n-d+1$ and $|J| = d-1$, respectively; cf. \cite{SchumacherWestmoreland:approx-QECC}.
Indeed, if Eq.~\eqref{eq:KLF_in_entropies} holds, then we can apply
the decoding isometry $V':X_I\rightarrow B_{f}F'$ in order to recover
the pure entangled state between $R$ and the first output register
$B_f$ of the isometry, whereas the other output register $F'$ is left
in a pure entangled state with $F X_J$.  Taking averages over
partitions, we get
\begin{equation}
  S(R) = \EE_{|I|=n-d+1} S(X_I) - \EE_{|J| = d-1} S(X_J).
\end{equation}
Here, $\EE_I S(X_I)$ is the average von Neumann entropy of subsystems with size $|I| = n-d+1$.
We now make use of Eq.~\eqref{eq:S-montononicity} from Lemma~\ref{lemma:entropy} below. 
Choosing $m=n-d+1$ and $\mu=d-1$, one has then
\begin{align}
  \label{eq:QMDS_purity_inequality}
  S(R) &=    \EE_{|I|=n-d+1} S(X_I) - \EE_{|J|=d-1} S(X_J) \notag\\
       &\leq \left(\frac{n-d+1}{d-1}-1\right) \EE_{|J|=d-1} S(X_J)\,.
\end{align}
This proves the claim. Note that we could apply the lemma because it is known
that any nontrivial code of distance $d$ must satisfy $d-1 < \frac{n}{2}$.
\end{IEEEproof}

\medskip
Note that the entropic Singleton bound \eqref{eq:entropic_bound} takes
the average entropy \eqref{eq:average entropy} of a random subblock of
size $d-1$ of the code into account. Thereby, it refines the bound
$\log_2 K \le (n-2d+2)\log_2 q$.

\begin{lemma}
\label{lemma:entropy}
Consider the $n$-party system $X_1 X_2\ldots X_n$, and for a subset 
$I\subset[n]$ of the ground set denote $X_I = \bigotimes_{i\in I} X_i$.
Let $\mu < m \leq n$. Then, with respect to any state $\rho$,
\begin{equation}
  \label{eq:S-montononicity}
  \EE_{|I|=m} S(X_I) \leq \frac{m}{\mu} \EE_{|J|=\mu} S(X_J) ,
\end{equation} 
where both expectation values are with respect to uniformly random
subsets $I,J\subset[n]$ of the ground set, of the respective cardinality.
\end{lemma} 

This lemma appears to have been stated first in \cite{Aharonov-et-al}, albeit with 
an incorrect proof; then again in \cite{JungePalazuelos}, with a proof attributed to AW;
see \cite{M-HFW} for a generalized perspective. For later use we will actually prove
directly a generalization of it to the conditional entropy: 

\begin{lemma}
\label{lemma:conditional-entropy}
Consider an $(n+1)$-party system $X_1 X_2\ldots X_n Y$ and denote $X_I = \bigotimes_{i\in I} X_i$ 
for any subset $I\subset[n]$. 
Let $\mu < m  \leq n$. Then, with respect to any state $\rho$,
\begin{equation}
  \EE_{|I|=m} S(X_I|Y) \leq \frac{m}{\mu} \EE_{|J|=\mu} S(X_J|Y) ,
\end{equation} 
where both expectation values are with respect to uniformly random
subsets $I,J\subset[n]$ of the respective cardinality.
\end{lemma}

\begin{IEEEproof}
We start with the case $n=m > \mu=n-1$. For this, we have to show that
\begin{alignat}{5}
  S(X_{[n]}|Y) &\leq \frac{n}{n-1} \EE_J S(X_J|Y)  \notag\\
  &= \frac{1}{n-1} \sum_{i=1}^n S(X_{[n]\setminus\{{i\}}}|Y). \label{eq:base-case-conditional}
\end{alignat}
Purifying the state to a pure state $\ket{\psi}^{X_{[n]}YZ}$ with an
auxiliary system $Z$, we can rewrite the left hand
side as $S(X_{[n]}|Y)=S(Z)-S(Y)$, and the terms in the sum on the right hand side as
$S(X_{[n]\setminus\{i\}}|Y) = S(X_iZ)-S(Y)$, making Eq.~\eqref{eq:base-case-conditional}
equivalent to 
\begin{equation}
  (n-1) S(Z) - (n-1) S(Y) \leq \sum_{i=1}^n S(X_iZ) - n S(Y).
\end{equation}
Adding $nS(Y)$ and subtracting $nS(Z)$ from both sides, this becomes equivalent to 
\begin{align}
S(Y)-S(Z)
&=    S(X_{[n]}Z)-S(Z) \nonumber\\
&=    S(X_{[n]}|Z) 
\leq 
\sum_{i=1}^n S(X_i|Z)\,,
\end{align}
which is indeed true by strong subadditivity. 

Repeated application of Eq.~\eqref{eq:base-case-conditional}
leads now, for each $I\subset [n]$ with $|I|=m$, to
\begin{equation}
  S(X_I|Y) \leq \frac{m}{\mu} \EE_{J\subset I,|J|=\mu} S(X_J|Y). \label{eq:cond_entropy_subset}
\end{equation}
Taking the average over all subsets $I$ concludes the proof.
\end{IEEEproof}

\medskip
A code that saturates the quantum Singleton bound from Theorem~\ref{thm:singleton} with equality
is called {\em quantum minimum distance separable (QMDS)}. 
Taking advantage of the above entropic proof, we can derive some interesting properties of such codes.
To state the following corollary, recall that a code of distance $d$ is called 
\emph{pure} if $\bra{\phi} E \ket{\phi} = 0$ holds for all code states $\ket{\phi}$ and for 
all traceless errors $E$ of weight smaller than $d$.
This is equivalent to every code state having maximally mixed $(d-1)$-body marginals.

\begin{corollary}[Cf.~\cite{HuberGrassl}]
\label{prop:QMDS_pure}
  Let $\mathcal{Q}=(\!(n,K,d)\!)_q$ be a QMDS code. Then it is pure. 
\end{corollary}
\begin{IEEEproof}
For a QMDS code, $S(R) =\log_2 K$.
Thus to satisfy Eq.~\eqref{eq:QMDS_purity_inequality}, 
$S(X_J) = (d-1)\log_2 q$ for all $J\subset[n]$ of size $d-1$
must hold. This proves the claim. 
\end{IEEEproof}

\medskip
Thus, we can make a statement about the entanglement that is necessary for QMDS codes to exist:  
Any code state necessarily needs to have the same $(d-1)$-marginals as the maximally mixed state 
on the code subspace. Consequently any code state of a QMDS code is maximally entangled across each 
$d-1$ vs. $n-d+1$ bipartition. 
Vice versa, if the code states are less entangled, it means that $\overline\sigma$ is less 
than its maximum value $\log_2 q$, and that additionally limits the amount of information 
that can be encoded.

\section{Quantum Singleton bound\protect\\ for general EAQECC}
\label{sec:EAQECC-Singleton}
We now show that the above ideas carry over to the analysis of 
EAQECCs. We start with a bound on the net entanglement generation of such
codes, which is most naturally discussed in the setting of CQECCs.

\begin{theorem}[Catalytic quantum Singleton bound]\label{thm:ea-singleton}
Consider a catalytic entanglement generation code as in Fig.~\ref{fig:leaf}
with net entanglement production $\ell_f-\ell_{in}$. 
Then, the latter is bounded by the quantum Singleton bound. 
Namely, with respect to the state $\ket{\psi}^{A_{f} F X_J X_I B_{in}}$ in
Fig.~\ref{fig:leaf}, one has
\begin{align}
  \ell_f - \ell_{in}   &=    S(A_{f}) - S(B_{in})  \nonumber\\
           &\leq \max\{0,(n-2d+2)\overline\sigma\}  \nonumber\\ &\leq \max\{0,n-2d+2\} \log_2 q \,.
\end{align}
Here, $S(B_{in})_\varphi = S(B_{in})_\psi$
and $S(A_{f})_{\varphi'} = S(A_{f})_\psi$
are the initial and final entanglement, respectively. As before, 
$\overline\sigma = \frac{1}{d-1}\EE_{|J|=d-1} S(X_J) \leq \log_2 q$
is the average entropy of a $(d-1)$-block per system.
\end{theorem}

\begin{IEEEproof}
By assumption, there is a 
decoding isometry $V'\colon X_I B_{in} \rightarrow B_{f} F'$ that
maps $\ket{\psi}^{A_{f}F X_I X_J B_{in}}$ to 
$\ket{\varphi'}^{A_{f}B_{f}} \otimes \ket{\omega}^{FF'X_J}$
with a suitable pure state $\omega$ of $FF'X_J$.
Thus,
\begin{alignat}{7}
  S(X_I B_{in})_\psi &= S(B_{f})_{\varphi'} + S(F')_\omega \notag\\
                          &= S(A_{f})_{\varphi'} + S(F X_J)_\omega\label{eq:product1}\\
                          &{}= S(A_{f})_{\psi} + S(F X_J)_\psi \notag\\
                          &{}= S(A_{f})_{\psi} + S(X_J)_\psi + S(F|X_J)_\psi,\label{eq:product2}
\end{alignat}
using the purity of the states $\varphi'$ and $\omega$.
On the other hand, by subadditivity of the von Neumann entropy, 
\[
  S(X_I B_{in})_\psi \leq S(X_I)_\psi + S(B_{in})_\psi,
\]
yielding
\begin{equation}
  \label{eq:entropy-entanglement}
  S(A_{f}) - S(B_{in}) \leq S(X_I) - S(X_J) - S(F|X_J).
\end{equation}
By taking the average over all partitions $[n] = I \stackrel{.}{\cup} J$ into blocks
of size $n-d+1$ and $d-1$, respectively, we get the basic bound
\begin{align}
  \label{eq:fundamental}
  S(A_{f}) - S(B_{in})  
      \leq \EE_{|I|=n-d+1} S(X_I) - \EE_{|J|=d-1} S(X_J) \notag\\ 
           - \EE_{|J|=d-1} S(F|X_J).
\end{align}

Let us start with the case $d-1 < n-d+1$.
We can then apply Lemma~\ref{lemma:entropy} with $m=n-d+1$ and $\mu=d-1$,
giving $\EE_{|I|=n-d+1} S(X_I) \leq \frac{n-d+1}{d-1} \EE_{|J|=d-1} S(X_J)$,
hence 
\begin{align}
  \ell_f - \ell_{in} &=     S(A_{f}) - S(B_{in}) \notag\\
                     &\leq \frac{n-2d+2}{d-1} \EE_{|J|=d-1} S(X_J) \notag\\
                     &\phantom{======:} - \EE_{|J|=d-1} S(F|X_J).
\end{align}
The first term on the r.h.s.~equals $(n-2d+2)\overline\sigma$, and we will
show that the other expectation value is non-negative. Indeed,
\begin{align}
  &\EE_{|J|=d-1} S(F|X_J)  \notag\\
        &\phantom{==}
         =    \frac12 \EE_{\substack{ |J|=|J'|=d-1 \\ J\cap J'=\emptyset }} 
                ( S(F|X_J)+S(F|X_{J'}) ) 
        \geq 0,
\end{align}
the latter because $S(F|X_J)+S(F|X_{J'}) \geq 0$ by strong subadditivity.

It remains to show $S(A_{f}) - S(B_{in}) \leq 0$ when $d-1 \geq n-d+1$.
Going back to Eq.~\eqref{eq:fundamental}, we reproduce the r.h.s.~by decomposing system $J$ as
$J = J_1 \stackrel{.}{\cup} J_2$, with $|J_1|=n-d+1$ and $|J_2|=2(d-1)-n$. 
We can then write
\begin{align}
  \EE_{|J|=d-1} \bigl(S(X_J)+S(F|X_J)\bigr) 
  &= \EE_{J_1,J_2} S(F X_{J_1} X_{J_2}) \notag\\
  &= \EE_{I,J_2} S(F X_{I} X_{J_2}) ,
\end{align}
and so we get
\begin{align}
  \ell_f - \ell_{in} &=    S(A_{f}) - S(B_{in})   \notag\\
                     &\leq - \EE_{I,{J_2}} S(FX_{J_2}|X_I) \notag\\
                     &= - \EE_{J_1,J_2} S(FX_{J_2}|X_{J_1}),
\end{align}
where the expectations are with respect to uniformly random partitions
$[n] = I \stackrel{.}{\cup} J_1 \stackrel{.}{\cup} J_2$ with $|I|=|J_1|=n-d+1$ and $|J_2|=2(d-1)-n$.
We can now again argue with strong subadditivity: $S(FX_{J_2}|X_I) + S(FX_{J_2}|X_{J_1}) \geq 0$,
and so the right hand side of the previous displayed equation is non-positive.
\end{IEEEproof}

Theorem~\ref{thm:ea-singleton} provides the quantum Singleton bound in the most 
general setting of catalytic entanglement generation and without restrictions on 
the encoder, apart from being given by a completely positive trace-preserving 
(cptp) map. 
Even in the non-assisted setting it thus goes beyond the original 
assumptions of Knill and Laflamme \cite{q-singleton}, and of Rains \cite{Rains}, 
who treated stabilizer and general subspace codes.

It is interesting to reflect on the status of the bound $d-1 <
\frac{n}{2}$, which holds in the non-assisted case due to no-cloning
(cf.~the original quantum Singleton bound), and obviously requires a
non-trivial code of dimension larger than $1$.  We note that this
bound is no longer true in the case of EAQECC as a counterexample
provided in
Refs.~\cite{Markus-Gegenbeispiel,Markus-Gegenbeispiel-Paper} shows.
Nevertheless, the above theorem still gives it a meaning, because it
states that $\log_K-\ell\leq 0$ holds when $d-1 \geq \frac{n}{2}$ and
thus the code cannot be used to generate any net entanglement.  The
analogous (and well-known) Shannon-theoretic statement is that not
only can the $50$--$50$ erasure channel not transmit any quantum
information, or for that matter generate entanglement, but it cannot
even be used to increase the amount of entanglement given any
pre-shared entanglement.

The bound for the generation of net entanglement is clearly the best
possible: If $d-1 \geq \frac{n}{2}$ then $\ell_f-\ell_{in}$ can reach
at most $0$. In this case we can always recover the initial
entanglement fully at the end by simply not touching it; if $d-1
< \frac{n}{2}$ on the other hand, it is known that for sufficiently
large $q$ there are QMDS codes saturating the bound when $k=\log_q
K=n-2d+2$, in which case there is no need for any entanglement
assistance.

\section{Transmission-entanglement tradeoff}
\label{sec:full-tradeoff}
The net entanglement generation is only part of the story, since the
bound of Theorem~\ref{thm:ea-singleton} conflates the qubits sent
through the code with the entanglement consumed by it.  To see why
that happens, consider the extreme case when unlimited entanglement
can be consumed, i.\,e.~$\ell$ is unbounded. We can then beat the
unassisted quantum Singleton bound by the following
entanglement-assisted scheme: using dense coding for each of the
$q$-ary quantum systems (requiring $n$ maximally entangled states of
Schmidt rank $q$) we can turn each of the $n$ quantum channels of
$\log_2 q$ qubits into classical channels of $\log_2 q^2$ classical
bits.  As $d-1$ of these systems undergo erasure, we can use a
classical MDS code for alphabet size $q^2$, encoding $(n-d+1)\log_2
q^2$ messages. These messages in turn are used to teleport $n-d+1$
$q$-ary quantum systems, requiring another $(n-d+1)\log_2 q$ ebits.
Thus, we get $k=(n-d+1)\log_2 q$, saturating the \emph{classical}
Singleton bound!  Note that the entanglement consumption attaining the
same code size can be reduced to $(d-1)\log_2 q$ ebits; see the
discussion after the main theorem of this section.

Next, we derive the inequalities for the full tradeoff between the 
transmitted quantum information $\log_2 K$ and the consumed entanglement $\ell$.
Define by 
\begin{align}
\overline\sigma &:= \frac{1}{d-1}\EE_{|J|=d-1} S(X_J) \leq \log_2 q \\
\overline{\overline\sigma} &:= \frac{1}{n-d+1}\EE_{|I|=n-d+1} S(X_I) \leq \log_2 q
\end{align}
the average entropies of a $(d-1)$-block and an $(n-d+1)$-block per
system respectively.  We have the following theorem on the qubit
vs. ebit (i.\,e., $\log_2 K$ vs. $\ell$) tradeoff.

\begin{theorem}[Qubit-ebit Singleton bound]
\label{thm:ea-full-singleton}\ 
Consider an EAQECC with minimum distance $d$ that encodes a
$K$-dimensional quantum system into $n$ $q$-ary quantum systems and
whose net entanglement consumption is $\ell = \ell_{in}-\ell_f$ ebits,
as illustrated in Fig.~\ref{fig:EAQECC}.
Then the following bounds hold.

If $d-1 < \frac{n}{2}$, then
\begin{align}
  \label{eq:q-singleton}
  \log_2 K &\leq (n-2d+2) \overline\sigma + \ell , \\
  \label{eq:Q_E}
  \log_2 K       &\leq (n-d+1) \overline{\overline\sigma} \leq (n-d+1) \overline\sigma.
\end{align}

If $d-1 \ge \frac{n}{2}$, then
\begin{align}
  \label{eq:q-singleton-larged}
  \log_2 K &\leq \ell , \\
  \label{eq:Q_E-large-d}
  \log_2 K        &\leq (n-d+1) \overline{\overline\sigma}, \\
  \label{eq:piecewise-linear}
  \log_2 K        &\leq \frac{n-d+1}{3d-3-n}\bigl( \ell + (2d-2-n) \overline{\overline\sigma} \bigr).
\end{align}
The performance of an arbitrary EAQECC is thus bounded as
\begin{align}
  \label{eq:q-singleton-ultimate}
  &\log_2 K \leq\max\{0,n-2d+2\} \log_2 q + \ell , \\
  \label{eq:Q_E-ultimate}
  &\log_2 K        \leq (n-d+1)\log_2 q, 
  \intertext{
  and if $d-1\ge \frac{n}{2}$, then}
  \label{eq:piecewise-linear-ultimate}
  &\log_2 K        \leq \frac{n-d+1}{3d-3-n}\bigl( \ell + (2d-2-n)\log_2 q \bigr).
\end{align}
\end{theorem}

\begin{figure*}[tbp]
  \centerline{
    \begin{tabular}{ccc}
    \includegraphics[width=0.25\hsize]{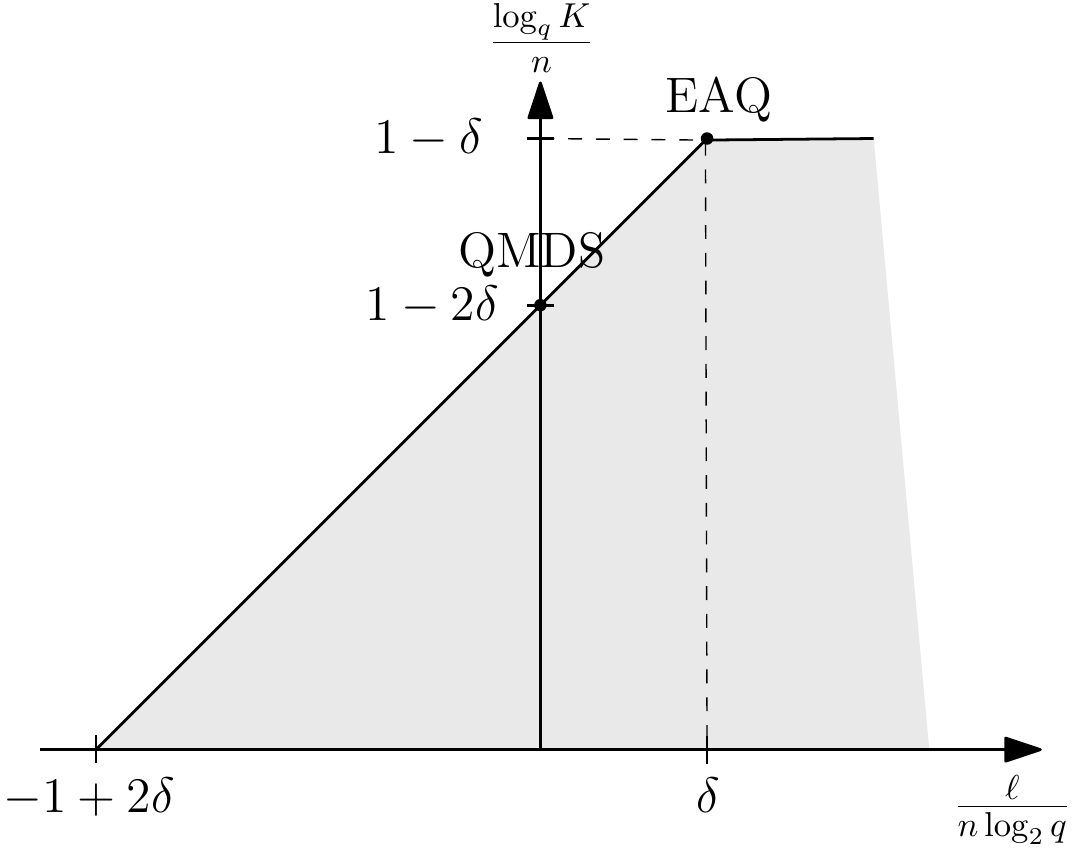}&
    \includegraphics[width=0.25\hsize]{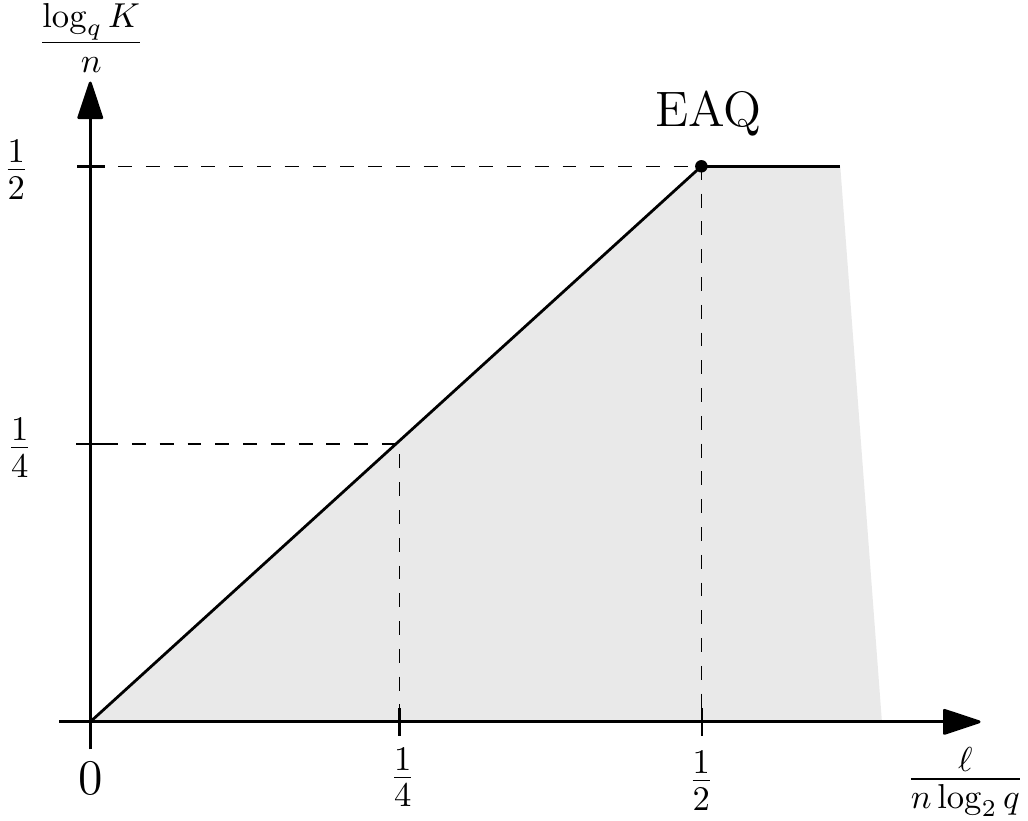}&
    \raisebox{-0.5pt}{\includegraphics[width=0.4\hsize]{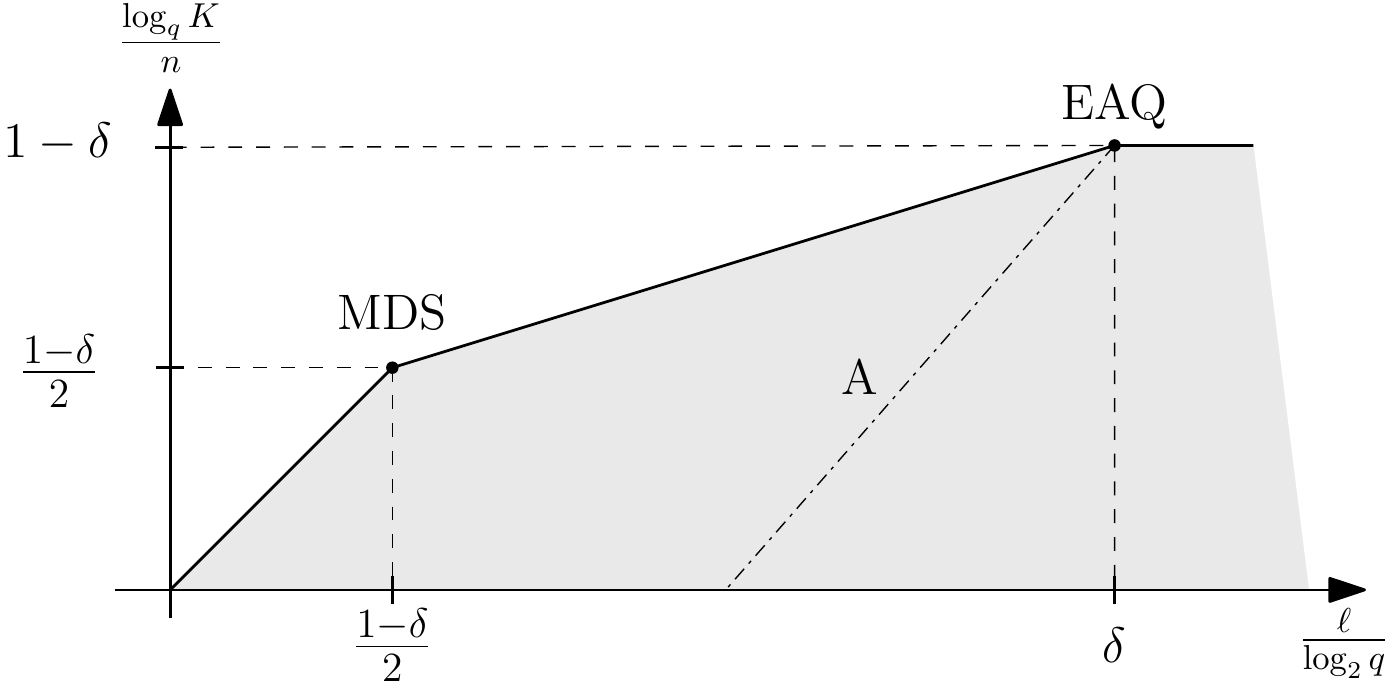}}\\
    {\footnotesize (a) $\delta=\frac{d-1}{n} < \frac{1}{2}$} &
    {\footnotesize (b) $\delta=\frac{1}{2}$} &
    {\footnotesize (c) $\delta>\frac{1}{2}$}
    \end{tabular}
  }
  \medskip
  
  \caption{The rate region showing the admissible pairs of $\log_2 K$ and $\ell$ (shown normalized
           with $n\log_2 q$), according to the minimum distance. There is a qualitative 
           difference between the cases $\delta = \frac{d-1}{n} < \frac12$ (a) and 
           $\delta=\frac12$ (b) on the one hand, and $\delta > \frac12$ (c) on the other.
           The maximum qubit rate is always $1-\delta$, attained with ebit rate $\delta$,
           in the former two cases that point (marked EAQ) contains the whole story, since
           moving along a line with slope $1$ to the left, converting qubits to ebits,
           we get to the maximum unassisted rate $1-2\delta$, and the maximum 
           entanglement generation rate $-(1-2\delta)$. (The negative sign is due to 
           our convention of interpreting $E$ as the net entanglement consumption, 
           which is a straightforward cost if $E\geq 0$, and generation of $-E$ ebits 
           of entanglement if $E<0$ \cite{QuShannonRT}.)
           In the latter case, the boundary develops a kink at $\frac{1-\delta}{2}$ (point MDS) and 
           becomes piecewise linear. The kink marks the change from Ineq.~\eqref{eq:Q_E-ultimate}
           being tight to Ineq.~\eqref{eq:piecewise-linear-ultimate} being tight.
           In the text we explain that the points EAQ and MDS are attained by 
           concrete codes, at least when $q$ is sufficiently large.
           The line $A$ marks a previous ``quantum Singleton bound''~\cite{BrunDevetakHsieh-catalytic}.}
  \label{fig:rate-region}
\end{figure*}

The shape of the rate region (the diagramme of admissible pairs
$\frac{\log_2 K}{n\log_2 q}=\frac{\log_q K}{n}$ and $\frac{\ell}{n\log_2 q}$)
is depicted in Fig.~\ref{fig:rate-region}. It depends on whether 
the normalized distance $\delta = \frac{d-1}{n}$ is smaller or larger than $\frac12$.

\begin{IEEEproof}
Eqs.~\eqref{eq:q-singleton} and \eqref{eq:q-singleton-larged} have already been shown 
in Theorem~\ref{thm:ea-singleton}.
The proofs of the three other Eqs.~\eqref{eq:Q_E}, \eqref{eq:Q_E-large-d},
and \eqref{eq:piecewise-linear} rest on entropic bounds that are
similar to those used for Theorem~\ref{thm:ea-singleton}.

Concretely, with respect to the pure state $\ket{\psi}^{RA_{f}F X_J
  X_I B_{in}}$, we claim that for a random partition $[n]=I
\stackrel{.}{\cup} J$ into sets of size $n-d+1$ and $d-1$,
respectively:
\begin{align}
  \log_2 K-\ell &\leq \EE_{|I|= n-d+1, |J| = d-1} \bigl( S(X_I)-S(X_J) \bigr) \notag\\ 
  &=: n\Delta. \label{eq:coherent-info-bound}
\end{align}
Furthermore, we claim that for any partition $[n]=I \stackrel{.}{\cup} J$,
\begin{align}
  \label{eq:mutual-information-bound}
  2\log_2 K &\leq I(RA_{f}B_{in}F:X_I) \notag\\
  &= S(X_{[n]})+S(X_I)-S(X_J) \notag\\
  &= S(X_I)+S(X_I|X_J) \leq 2 S(X_I)\,.
\end{align}

To prove Eq.~\eqref{eq:coherent-info-bound}, we first do so assuming trivial $F$.
For this, it is enough to consider the situation as in Fig.~\ref{fig:leaf},
which we have already done in the proof of Theorem~\ref{thm:ea-singleton}.
Namely, Eq.~\eqref{eq:entropy-entanglement}, with trivial $F$ system,
gives precisely 
\[
  \log_2 K-\ell = S(A_{f}) - S(B_{in}) \leq S(X_I) - S(X_J), 
\]
and by averaging over partitions we get, as desired,
\[
\begin{split}
  \log_2 K-\ell &=    S(A_{f}) - S(B_{in})  \\
                &\leq \EE_{|I|=n-d+1} S(X_I) - \EE_{|J|=d-1} S(X_J) .
\end{split}
\]

To prove Eq.~\eqref{eq:mutual-information-bound}, we use the data
processing inequality repeatedly, as well as the fact that $R$ and
$B_{in}$ are decoupled, $I(R:B_{in})=0$. Then
\[\begin{split}
  2\log_2 K &=    I(R:\widehat{M}) \\
            &\leq I(R:B_{in}X_I) \\
            &=    I(R:B_{in}) + I(R:X_I|B_{in}) \\
            &\leq I(RB_{in}:X_I) \\
            &\leq I(RB_{in}A_{f}F:X_I) .
\end{split}\]

This allows to prove Eqs.~\eqref{eq:Q_E}, \eqref{eq:Q_E-large-d}, and
\eqref{eq:piecewise-linear}. Namely, averaging
Eqs.~\eqref{eq:coherent-info-bound} and
\eqref{eq:mutual-information-bound} over partitions yields
\begin{align}
  \label{eq:Delta}
  2\log_2 K &\leq S(X_{[n]}) + n\Delta, \\
   \log_2 K &\leq \EE_{|I|= n-d+1} S(X_I) =
   (n-d+1)\overline{\overline\sigma}.\label{eq:average_S_X_I}
\end{align}
Now, Eq.~\eqref{eq:Q_E} follows from \eqref{eq:average_S_X_I},
invoking once more Lemma~\ref{lemma:entropy}; Eq.~\eqref{eq:Q_E-large-d} follows
from Eq.~\eqref{eq:average_S_X_I}, too. To proof
Eq.~\eqref{eq:piecewise-linear}, we proceed as follows:
\[\begin{split}
  S(X_{[n]})-S(X_I) &=    S(X_J|X_I) \\
                &\leq \frac{d-1}{2d-2-n}\EE_{|J'|=2d-2-n} S(X_{J'}|X_I),
\end{split}\]
using Eq.~\eqref{eq:cond_entropy_subset}  with $J'\subset J$,
$|J|=d-1$ and $|J'|=d-1-(n-d+1)=2d-2-n$. Taking the average over $I$ as well, we get
\[\begin{split}
  &S(X_{[n]}) \leq \EE_I S(X_I) + \frac{d-1}{2d-2-n}\EE_{I,J'} S(X_{J'}|X_I) \\
         &\phantom{==}
          =    (n-d+1)\overline{\overline\sigma} 
                 + \frac{d-1}{2d-2-n}\bigl( \EE_J S(X_J) - \EE_I S(X_I) \bigr) \\
         &\phantom{==}
          =    (n-d+1)\overline{\overline\sigma} - \frac{d-1}{2d-2-n}n\Delta. 
\end{split}\]
Plugging this into Eq.~\eqref{eq:Delta}, we obtain
\[\begin{split}
  2\log_2 K &\leq S(X_{[n]}) + n\Delta \\
            &\leq (n-d+1)\overline{\overline\sigma} - \left( -1 + \frac{d-1}{2d-2-n}\right) n\Delta \\
            &=    (n-d+1)\overline{\overline\sigma} - \frac{n-d+1}{2d-2-n} n\Delta \\
            &\leq (n-d+1)\overline{\overline\sigma} - \frac{n-d+1}{2d-2-n}(\log_2 K-\ell), 
\end{split}\]
where the last line follows from Eq.~\eqref{eq:coherent-info-bound}. 
Solving for $\log_2 K$ and simplifying leads to Eq.~\eqref{eq:piecewise-linear}.

It remains to show that the same bounds hold for general encoders, even with 
nontrivial $F$ system. We do so by reducing it to the case of trivial $F$.
Namely, consider any rank-one measurement $\{M_t\}$ on $F$, and define
measurement probabilities $p_t$ and post-measurement states by letting
$p_t \psi_t^{RA_{f}X_JX_IB_{in}} := \tr_F
\psi^{RA_{f}FX_JX_IB_{in}}(M_t^F \ox \1)$ (Note that
$\psi_t=\ket{\psi_t}\bra{\psi_t}$ denotes the density matrix of the
pure state $\ket{\psi_t}$).  Ignoring the measurement outcome $t$, we
obtain the following ensemble of pure states
\begin{equation}
  \label{eq:convex-t}
  \psi^{RA_{f}X_JX_IB_{in}} = \sum_t p_t \psi_t^{RA_{f}X_JX_IB_{in}}.
\end{equation}
The decoder $V'$ will work for all states $\psi_t$, resulting in the same
pure states $\Phi_K^{R\widehat{M}}$ (maximally entangled) and
${\varphi'}^{A_{f}B_{f}}$: indeed, $I(F:RA_{f})=0$.  In other words,
$F$ is decoupled from $RA_{f}$, which means that conditional on $t$,
the state of $RA_{f}$ is the same as the original $\psi^{RA_{f}}$.

An important consequence of this is that $\psi_t^R = \psi^R = \frac{1}{K}\1_R$ is
the maximally mixed state for all~$t$. 
This in turn implies that there exists states $\ket{\varphi_t}^{A_{in}B_{in}}$
and isometries $V_t\colon MA_{in} \hookrightarrow X_{[n]} A_{f}$ such that
\[
  \ket{\psi_t}^{RA_{f}X_JX_IB_{in}}
       = (V_t\ox\1_{RB_{in}})(\ket{\Phi_K}^{RM}\ket{\varphi_t}^{A_{in}B_{in}}).
\]
In other words, we have got an EAQEC for each $t$, consisting of the
encoding and decoding isometries $V_t$ and $V'$, respectively, and with 
initial and final entangled states $\varphi_t$ and $\varphi'$, respectively.
For a given $t$, such a code has the following properties: 
it transmits $\log_2 K$ qubits as 
the original code, and it has a net entanglement consumption of
$\ell_t = S(B_{in})_{\varphi_t} - S(A_{f})_{\varphi'}
        = S(B_{in})_{\psi_t} - S(A_{f})_{\psi_t}$.
By Eq.~\eqref{eq:convex-t}, we have $\sum_t p_t \ell_t \leq \ell$, 
where we used concavity of the von Neumann entropy 
(note that $t$ only affects the initial,
but not the final entanglement).

At the same time, the above proof shows that for each $t$, we have the 
bounds \eqref{eq:q-singleton}--\eqref{eq:piecewise-linear} for $\log_2 K$ and $\ell_t$,
with average entropies ${\overline\sigma}_t = \frac{1}{d-1} \EE_{|J|=d-1} S(X_J)_{\psi_t}$
and ${\overline{\overline\sigma}}_t = \frac{1}{n-d+a} \EE_{|I|=n-d+1} S(X_I)_{\psi_t}$
on the right hand side. Once more by Eq.~\eqref{eq:convex-t} and concavity of
the entropy, we have $\sum_t p_t {\overline\sigma}_t \leq {\overline\sigma}$
and $\sum_t p_t {\overline{\overline\sigma}}_t \leq \overline{\overline\sigma}$,
concluding the proof.
\end{IEEEproof}

A code that has extremal parameters with respect to the quantum
Singleton bound of
Eqs.~\eqref{eq:q-singleton-ultimate}--\eqref{eq:piecewise-linear-ultimate}
in Theorem~\ref{thm:ea-full-singleton} is called {\em
  entanglement-assisted quantum minimum distance separable
  (EAQMDS)}.  The parameters of these codes are on the upper boundary
of the regions in Fig.~\ref{fig:rate-region}~(a)--(c).

We recall that for codes with no entanglement assistance the bound
$d<\frac{n}{2}+1$ holds due to a no-cloning argument.
Eq.~\eqref{eq:q-singleton-ultimate} then yields the usual quantum
Singleton bound for QECC, $\log_2 K \leq (n - 2d +2)\log_2 q$.  As
shown in Corollary~\ref{prop:QMDS_pure}, a QMDS code meeting this
bound with equality must be pure. Next we show that this is no longer
the case for EAQMDS codes.  Recall that a quantum code is pure if all
$(d-1)$-body marginals are maximally mixed.  We make the assumption
that the auxiliary output system $F$ of the encoder $V_{\text{ENC}}$
in Fig.~\ref{fig:EAQECC} is trivial, as we have seen in the proof of
Theorem~\ref{thm:ea-full-singleton} that the bound is independent of
the dimension of $F$.  We include, however, the situation of catalytic
codes, i.\,e., the code might produce entanglement between the output
$A_f$ and $B_f$.

\begin{theorem}[Singleton bound for pure codes]
\label{thm:pure-singleton}\hfil 
Consider a pure EAQECC with minimum distance $d$ that encodes a
$K$-dimensional quantum system into $n$ $q$-ary quantum systems and
whose net entanglement consumption is $\ell$ ebits.  Then 
\begin{alignat}{5}
  \log_2 K \le (n-2d+2)\log_2 q+\ell.
\end{alignat}
\end{theorem}
\begin{IEEEproof}
  Consider the state $\ket{\psi}^{RA_fF X_JX_IB_{in}}$ in
  Fig.~\ref{fig:EAQECC}.  As we assume that the auxiliary output
  system $F$ is trivial, it can be omitted.  Similar as in the
  derivation of Eqs. \eqref{eq:product1} and \eqref{eq:product2} in
  the proof of Theorem \ref{thm:ea-singleton}, the existence of a
  decoding isometry $V'_{\text{DEC}}$ implies that
  \begin{alignat*}{5}
    S(X_IB_{in}) = S(X_J) + S(A_f) + S(R).
  \end{alignat*}
  Hence
  \begin{alignat}{5}
    \log_2 K &= S(R) = S(X_I B_{in})- S(X_J) -S(A_f) \notag\\
                    &\le S(X_I) + S(B_{in}) -S(A_f) - S(X_J) \notag\\
                    &\le (n-d+1)\log_2 q + \ell - (d-1)\log_2 q.
  \end{alignat}
  Here we have used subadditivity and the fact that for pure codes the
  reduced state on the subsystem $X_J$ is maximally mixed.
\end{IEEEproof}

\medskip
We note that a related result for the special case of non-degenerate
stabilizer codes has recently been derived in \cite[Thm.~10]{NadkarniGarani-2021}.

\begin{corollary}
  An EAQMDS code $\mathcal{Q}$ with $d\ge\frac{n}{2}+1$ with
  net entanglement consumption $0<\ell<(d-1)\log_2 q$ is not pure.
\end{corollary}
\begin{IEEEproof}
  For $d\ge\frac{n}{2}+1$, the parameters of an EAQMDS code are on the
  upper boundary of the region in Fig.~\ref{fig:rate-region}~(c).  By
  Theorem~\ref{thm:pure-singleton}, for pure codes the bound indicated
  by the line $A$ applies.  The bound for pure codes only agrees with
  the general bound for $\ell=0$ and at the point marked EAQ with
  $\ell=(d-1)\log_2 q$.
\end{IEEEproof}

\section{Optimal constructions}
\label{sec:constructions}
We now turn to a discussion of what we know about the tightness of the
various bounds derived in the preceding sections, depending on whether
$d-1 < \frac{n}{2}$ or $d-1 \ge \frac{n}{2}$.  For this purpose we use
the coding theory convention to read the logarithm in the formulas to
base $q$, and we will be dealing with codes of $n$ qudits (systems of
dimension $q$), and encoding a space of dimension $K=q^k$.  We use the
usual notation $(\!(n,K,d)\!)_q$ for a code of dimension $K$ with
minimum distance at least $d$ on $n$ $q$-ary systems; when $K=q^k$,
this is equivalent to $[\![n,k,d]\!]_q$.  For an EAQECC with code
dimension $K=q^k$ and minimum distance $d$ that has a net consumption
of $c=c_{in}-c_{f}$ maximally entangled pairs of qudits, we use the
notation $[\![n,k,d;c]\!]_q$ (see \cite{BrunDevetakHsieh-EA}). Note
that the net entanglement consumption is $\ell=c\log_2 q$ ebits.  For
the convenience of the reader, we rephrase Theorem
\ref{thm:ea-full-singleton} in this coding theoretic context.
\begin{corollary}
  Consider an EAQECC $[\![n,k,d;c]\!]_q$ with minimum distance $d$
  that encodes $k$ qudits into $n$ qudits with a net consumption of $c$
  maximally entangled pairs of qudits. Then the following holds:
  \begin{alignat}{5}
      k &{} \le c+\max\{0,n-2d+2\},\label{eq:q-singleton-ultimate2}\\
      k &{} \le n-d+1,\label{eq:Q_E-ultimate2}\\
      k &{} \le \frac{(n-d+1)(c+2d-2-n)}{3d-3-n}\quad\text{if $d-1\ge \frac{n}{2}$.}\label{eq:piecewise-linear-ultimate2}
  \end{alignat}
\end{corollary}

For all block lengths~$n$ and $d-1 \leq \frac{n}{2}$, there exist QMDS
codes with the parameters $[\![n,n-2d+2,d]\!]_q$ if only the alphabet
size $q$ is chosen large enough (see e.\,g.~Ref.~\cite{HuberGrassl} for
an overview). Naturally, then $c=0$, and in fact $c_{in}=c_{f}=0$.
This construction achieves the point QMDS in
Fig.~\ref{fig:rate-region}~(a).

We note that for $d-1<\frac{n}{2}$, using linear programming the bound
\eqref{eq:q-singleton-ultimate2} has been derived in
\cite{LaiAshikhmin} for qubits; the generalization to prime powers $q$
is presented in \cite{Allahmadi:2021}.

Consider now the general case, including that of $d-1 > \frac{n}{2}$.
For all $n$, $d$, and large enough $q$, there exist
entanglement-assisted quantum MDS (EAQMDS) codes with parameters
$[\![n,n-d+1,d;d-1]\!]_q$; here $c_{in}=d-1$ and $c_f=0$.  This
follows from the straightforward generalization
of~\cite[Cor.~2]{WildeBrun-2008} to prime power alphabets and the fact
that for alphabet size $q^2\ge 9$, any linear code is equivalent to a
code that trivially intersects its Hermitian dual code
\cite{Carlet-2018}.  These codes correspond to the point marked EAQ in
Fig.~\ref{fig:rate-region}~(a)--(c). The line descending from that
point to the lower left at slope $1$ in Fig.~\ref{fig:rate-region} (a)
and (b) is achievable by decreasing $k$ and simultaneously increasing
$c_{f}$ by the same amount.  In other words, we have the propagation
rule $[\![n,k,d;c]\!]_q \longrightarrow [\![n,k-1,d;c-1]\!]_q$.  To
see this, use the $[\![n,k,d;c]\!]_q$ code to encode $k-1$ qudits as
well as a maximally entangled state $\ket{\Phi_q}$; its decoder will
recover the $k-1$ qudits, $c_f$ maximally pairs of maximally entangled
pairs of qudits, plus another maximally entangled state
$\ket{\Phi_q}$. This additional maximally entangled pair of qudits
reduces the net consumption of maximally entangled pairs to $c-1$.

Somewhat surprisingly, any pure quantum code can be used to construct an EAQECC. 
\begin{theorem}
  \label{thm:pureQECC_EAQECC}\ {}
  Let $\mathcal{Q}$ be a pure QECC with parameters $[\![n,k,d]\!]_q$.
  Then EAQECC codes with parameters ${[\![n-c,k,d;c]\!]_q}$ exist for all $c<d$.
\end{theorem}

\begin{IEEEproof}
Assume that we have a pure code $\mathcal{Q}=(\!(n,K,d)\!)_q$ with an encoding
isometry $V\colon M\rightarrow A_{f}FX_{[n]}$. We input
half of a maximally entangled state $\ket{\Phi_K}^{RM}$ and obtain a
pure state on $R A_{f}F X_{[n]}$. We split $X_{[n]}$ into
two systems $X_{[n-c]}$ and $B_{in}$ of size $n-c$ and
$c<d$, respectively.  As the code is pure and has minimum distance
$d>c$, the reduced state on $B_{in}$ is maximally mixed.
Consider the Schmidt decomposition
\begin{alignat}{5}
V^M\ket{\Phi_K}^{RM}
  &= \ket{\psi}^{RA_{f}FX_{[n-c]B_{in}}} \nonumber\\
  &= \frac{1}{\sqrt{q^c}}\sum_{i=1}^{q^c}\ket{\psi_i}^{RA_{f}F X_{[n-c]}} \ket{\widetilde{\psi}_i}^{B_{in}},
\end{alignat}
where the states $\ket{\widetilde{\psi_i}}$ form an orthonormal basis
of the system $B_{in}$.  Hence there is a unitary transformation $U$ on $B_{in}$
that maps the standard basis on $B_{in}$ to this basis.
Similarly, there is an isometry 
$W\colon M A_{in}\rightarrow A_{f}F X_{[n-c]}$
such that
\begin{align}
&\left(W^{M A_{in}}\otimes U^{B_{in}}\right)
\left(\ket{\Phi_K}^{RM}\ket{\varphi}^{A_{in}B_{in}}\right)\notag\\
&\qquad=\frac{1}{\sqrt{q^c}}\sum_{i=1}^{q^c}\ket{\psi_i}^{RA_{f}F X_{[n-c]}} \ket{\widetilde{\psi}_i}^{B_{in}},
\end{align}
where $\ket{\varphi}^{A_{in}B_{in}}$ is a
maximally entangled state.

As $U$ acts only on $B_{in}$, it can be applied by the
receiver, followed by the decoding isometry for the original code.
That shows that the minimum distance of the new code is at least $d$.
\end{IEEEproof}

This propagation rule generalizes that from Ref.~\cite{LaiBrun2012}
for pure qubit stabilizer codes to arbitrary pure codes, and corrects
and generalizes the original propagation rule from
Ref.~\cite{Galindo2019} that would violate the EAQEC Singleton bound
[Theorem~\ref{thm:ea-full-singleton}] when starting with a stabilizer
QMDS code (see also Ref.~\cite{Galindo2021}).

\begin{corollary}\label{cor:prop_rule}
Any QMDS code with parameters $[\![n,n-2d+2, d]\!]_q$
gives rise to an EAQMDS code with the parameters $[\![n-c, n-2d+2, d; c]\!]_q$ for all $c<d$.
\end{corollary}

\begin{IEEEproof}
Proposition~\ref{prop:QMDS_pure} states that all QMDS codes are pure. 
The claim follows from Theorem~\ref{thm:pureQECC_EAQECC}.
\end{IEEEproof}

This construction works in particular when starting from an absolutely maximally 
entangled (AME) state with an even number $n$ of parties. 
These are pure quantum states for which maximal entanglement is present across 
every bipartition.
As with QMDS codes, such states always exist as long as the local dimension $q$ 
is chosen large enough.
(For example, the CSS construction of a so-called Euclidean QMDS code requires 
$n\leq q+1$ for $2<q$.)
For an even number $n$ of parties, 
AME states are QMDS codes with parameters $[\![n,0,\frac{n}{2}+1]\!]_q$.
With the propagation rule for unassisted pure codes 
($[\![n,k,d]\!]_q \longrightarrow [\![n-1,k+1,d-1]\!]_q$, 
c.f. Refs.~\cite{Rains2} and \cite{HuberGrassl})
one obtains $[\![n-k,k,\frac{n}{2}+1-k]\!]_q$ QMDS codes.
With Corollary~\ref{cor:prop_rule}
this yields EAQMDS codes with parameters $[\![n-k-t,k,\frac{n}{2}+1-k;t]\!]_q$
for $n+2=k+2d$ and $\frac{n}{2}+1-k > t$. 
(See \cite{HuberGrassl} for the 
discussion of other propagation rules within the class of unassisted QECC.)
This shows once more that the point EAQ in Fig.~\ref{fig:rate-region}~(a)--(c)
is attained.
Nevertheless, the whole set of QMDS codes is in a sense strictly more
powerful in this construction than when one considers solely AME
states.  As an example, a QMDS code with parameters $[\![8,4,3]\!]_3$
can be used to yield a $[\![7,4,3;1]\!]_3$ EAQMDS code.  This cannot
be realized with AME states: it has been shown that there are no AME
states on $2n-2t = 14-2 = 12$~qutrits, in other words, a
$[\![12,0,7]\!]_3$ code does not exist~\cite{HuberGrassl}.

Continuing with the case $d-1 \ge \frac{n}{2}$, we turn our attention 
to the point marked MDS in Fig.~\ref{fig:rate-region} (c)
which has $k=c=\frac{1}{2}(n-d+1)$.
It is also achievable:
use a classical MDS code to encode $(n-d+1)\log_2 q$ bits on $n$ quantum systems. 
These bits are used to teleport $\frac12(n-d+1)\log_2 q$
qubits, using $\frac12(n-d+1)\log_2 q$ ebits in entanglement assistance. 
This is the basic idea underlying the counterexamples in Ref.~\cite{Markus-Gegenbeispiel}. 
By simply using less information from the classical MDS code, and
concomitantly less entanglement, we can clearly attain the entire straight 
line connecting the point MDS to the origin in (c). 

It is unclear what the status of the line connecting the points MDS
and EAQ in Fig.~\ref{fig:rate-region}~(c) is; however, one might
conjecture that it is ``essentially'' attainable for large enough
alphabet size $q$, i.\,e.~all integer points below it correspond to
possible code values. This would for example follow if there exist
hybrid codes that transmit certain numbers of classical bits and
qubits assisted by entanglement, after which we convert the classical
bits to qubits by consuming even more entanglement in teleportation.
Concretely, it would be enough to find EAQECC $[\![n,k,d;c]\!]_q$ with
$k = \frac{n-d+1}{d-1}c$ that can \emph{simultaneously} transmit $j =
\frac{n-d+1}{d-1}(d-1-c)$ classical bits.  Note that in this case,
$j+k=n-d+1$, which is the relation for the parameters of a classical
MDS code.

What about other propagation rules, for example is there a chance to turn any 
$[\![n,k+c,d;c]\!]_q$ into an $[\![n,k,d]\!]_q$?
This does not seem to be the case in general: 
for example, a $[\![4,1,3;1]\!]_2$ EAQMDS code has been constructed~\cite{BrunDevetakHsieh-EA}, 
while it has been shown in Ref.~\cite{HiguchiSudbery2000} that no $[\![4,0,3]\!]_2$ exists. 
It is worth pointing out that the propagation rule described at the beginning 
of this section yields a $[\![4,0,3;0]\!]_2$ EAQECC, which is no contradiction, 
as it has $c_{in}=c_f=1$ rather than $0$.

\section{Discussion}
\label{sec:disc}
The bounds from Theorem~\ref{thm:ea-full-singleton} give precisely the 
qubit-ebit capacity region of the erasure channel, for erasure probability 
$\delta \in [0,1]$, and for any alphabet size $q$ \cite{HsiehWilde}. 
(We remark that the treatment in \cite{HsiehWilde}
seems implicitly to have assumed $\delta < \frac12$, though.)

In particular, the region is not just an outer bound but several lines and 
isolated points of it are also attainable by codes of sufficiently large alphabet
size $q$. 
This is no coincidence, as the Singleton bound really is about the capability of 
a code to correct erasure errors. Furthermore, the consistent use of the 
von Neumann entropy in the proofs makes the bounds robust to small 
deviations from the ideal. That is, they will essentially still hold even 
if the decoder works only for most of the erasure pattern subsets $J$
of size $d-1$, and only decodes with fidelity close to $1$. This is because 
the decoupling condition for QECC will be replaced by a trace distance bound of 
$\epsilon$ between the actual state and the decoupled state \cite{SchumacherWestmoreland:approx-QECC}, 
and this results in Eq.~\eqref{eq:KLF_in_entropies}, and analogously 
in later proofs, being replaced by an approximate equality, up to terms 
$O(\epsilon n\log_2 q+h(\epsilon))$, by the Fannes continuity inequality for the 
von Neumann entropy and conditional entropy \cite{Fannes,Audenaert,Winter:Fannes}. 
The rest of the proof follows unchanged. 

There are many open problems left, the first being the attainability of the
bounds along the line MDS-EAQ when $d-1>\frac{n}{2}$, 
at least for sufficiently large alphabet size $q$. We know that the endpoints
are attained in this way, and so we are asking for some sort of interpolation 
between the two corresponding constructions. Note that our bounds cut out a 
convex region, but a priori there is no reason why the actually achievable
codes and their qubit-ebit tradeoff should have any sort of convexity property. 
Returning briefly to the Shannon theoretic model of only demanding high-fidelity 
decoding for most erasure patterns, though, we observe that there at least the 
convexity is essentially granted: This is because we can juxtapose codes 
of block length $n_1$ and of block length $n_2$ for the same relative distance 
$\delta$, and get one that corrects with high fidelity for most erasures, 
at marginally smaller relative distance $\delta-\epsilon$ of block length $n_1+n_2$, 
while the code lengths add. 

A second important question is, what are the special properties characterizing 
the codes attaining the bounds?

Third, what are the constraints on the alphabet size $q$ for an EAQMDS code 
$[\![n,k,d;c]\!]_q$ to exist?
For QMDS codes is it known that for $d \geq 3$ the alphabet size must satisfy 
$n \leq q^2+d-2$~\cite{HuberGrassl}, and it is reasonable to expect that a similar scaling 
should hold for EAQMDS.

Finally, how do these bounds compare to previous algebraic quantum Singleton-type bounds,
and in particular to the recent results of Lai and Ashikhmin \cite{LaiAshikhmin} 
for EAQECC?

\section*{Acknowledgments}
We thank Andrew Nemec as well as the anonymous reviewers for valuable
comments on earlier versions of the manuscript.  
Questions by one of the latter led to the formulation of Theorem
\ref{thm:pure-singleton}.

MG acknowledges support by the Foundation for Polish Science
(IRAP project, ICTQT, contract no. 2018/MAB/5, co-financed by EU
within Smart Growth Operational Programme).

FH was supported by the Fundaci\'o Cellex,
the Spanish MINECO (projects QIBEQI FIS2016-80773-P and Severo Ochoa SEV-2015-0522),
the Generalitat de Catalunya (SGR-1381 and CERCA Programme), 
the European Union under Horizon2020 (PROBIST 754510), 
and the Foundation for Polish Science through TEAM-NET
(POIR.04.04.00-00-17C1/18-00).

AW acknowledges financial support 
by the Spanish MINECO (projects FIS2016-86681-P
and PID2019-107609GB-I00/AEI/10.13039/501100011033) 
with the support of FEDER funds, 
and the Generalitat de Catalunya (project 2017-SGR-1127).

\begin{IEEEbiographynophoto}{Markus Grassl}
(S'94--M'00--SM'20) received his diploma degree in
Computer Science in 1994 and his doctoral degree in 2001, both from
the Fakult\"at f\"ur Informatik, Universit\"at Karlsruhe (TH),
Germany. From 1994 to 2007 he was a member of the Institut f\"ur
Algorithmen und Kognitive Systeme, Fakult\"at f\"ur Informatik,
Universit\"at Karlsruhe (TH), Germany.

From 2007 to 2008 he was with the Institute for Quantum Optics and
Quantum Information of the Austrian Academy of Sciences in
Innsbruck. From 2009 to 2014, he was a Senior Research Fellow at the
Centre for Quantum Technologies at the National University of
Singapore. In 2014, he joined the Friedrich-Alexander-Universit\"at
Erlangen-N\"urnberg and the Max Planck Institute for the Science of
Light (MPL), Erlangen. Since 2019, he is a Senior Scientist at the
International Centre for Theory of Quantum Technologies, University of
Gdansk.  His research interests include quantum computation, focusing
on quantum error-correcting codes, and methods of computer algebra in
algebraic coding theory.  He maintains tables of good block quantum
error-correcting codes as well as good linear block codes.

Dr. Grassl served as Associate Editor for Quantum Information Theory of the IEEE
Transactions on Information Theory from 2015 till 2017.
\end{IEEEbiographynophoto}

\vfill\pagebreak

\begin{IEEEbiographynophoto}{Felix Huber} received a M.S. degree in physics
from ETH Z\"urich, Switzerland, in 2012, 
and a Ph.D. degree in theoretical quantum optics from the Universit\"at Siegen, Germany, in 2017.

In 2018 he was a Postdoctoral Researcher at the Universit\"at zu K\"oln, Germany, 
and from 2018 to 2020 he was a Research Fellow at ICFO Barcelona, Spain.
Since 2021 he is an Adiunkt at the Uniwersytet Jagiello\'nski in Krak\'ow, Poland.
His research interests include quantum error correction, multipartite entanglement, 
and invariant theory.

Dr. Huber is recipient of the 2019 Dissertation Prize by the 
section AMOP of the German Physical Society. 
\end{IEEEbiographynophoto}

\begin{IEEEbiographynophoto}{Andreas Winter} received a Diploma degree 
in Mathematics from the Freie Universit\"at Berlin, Germany, in 1997, 
and a Ph.D. degree from the Fakult\"at f\"ur Mathematik, Universit\"at 
Bielefeld, Germany, in 1999.

He was Research Associate at the University of Bielefeld until 2001,
and then with the Department of Computer Science at the University of
Bristol, UK. In 2003, still with the University of Bristol, he was
appointed Lecturer in Mathematics, and in 2006 Professor of Physics of
Information. Since 2012 he has been ICREA Research Professor with the
Universitat Aut\`onoma de Barcelona, Spain. His research interests
include quantum and classical Shannon theory, and discrete
mathematics.

Prof. Winter is recipient, along with Charles H. Bennett, Igor Devetak, 
Aram W. Harrow and Peter W. Shor, of the 2017 Information Theory Society Paper Award.
\end{IEEEbiographynophoto}

\vfill
\phantom{.}


\begin{thebibliography}{99}
\bibitem{GrasslBethPellizarri1997} Markus Grassl, Thomas Beth, and Thomas Pellizzari,
  ``Codes for the quantum erasure channel'',
  \emph{Phys. Rev. A} {\bf 56}(1):33--38, 1997.

\bibitem{singleton} Richard C. Singleton, ``Maximum distance $q$-nary codes'', 
  \emph{IEEE Trans. Inf. Theory} {\bf 10}(2):116--118, 1964. 

\bibitem{q-singleton} Emanuel Knill and Raymond Laflamme,
  ``A theory of quantum error-correcting codes'', 
  \emph{Phys. Rev. A} {\bf 55}(2):900--911, 1997.

\bibitem{Rains} Eric M. Rains, ``Nonbinary Quantum Codes'',
  \emph{IEEE Trans. Inf. Theory} {\bf 45}(6):1827--1832, 1999. 

\bibitem{Klappi} Andreas Klappenecker and Pradeep K. Sarvepalli, 
  ``On subsystem codes beating the quantum Hamming or Singleton bound'',
  \emph{Proc. Roy. Soc. London A} {\bf 463}(2087):2887--2905, 2007.

\bibitem{BDSW} Charles H. Bennett, David P. DiVincenzo, John A. Smolin, and 
  William K. Wootters, 
  ``Mixed-state entanglement and quantum error correction'', 
  \emph{Phys. Rev. A} {\bf 54}(5):3824--3851, 1996. 
  
\bibitem{HuberGrassl} Felix Huber and Markus Grassl, 
 ``Quantum Codes of Maximal Distance and Highly Entangled Subspaces'',
  \emph{Quantum} {\bf 4}:284, 2020.

\bibitem{BrunDevetakHsieh-EA} Todd A. Brun, Igor Devetak and Min-Hsiu Hsieh, 
  ``Correcting quantum errors with entanglement'',
  \emph{Science} {\bf 314}(5798):436--439, 2006. 

\bibitem{BrunDevetakHsieh-catalytic} Todd A. Brun, Igor Devetak and Min-Hsiu Hsieh,
  ``Catalytic quantum error correction'', 
  \emph{IEEE Trans. Inf. Theory} {\bf 60}(6):3073--3089, 2014.
 
  
\bibitem{SchumacherWestmoreland:approx-QECC} Benjamin Schumacher and Michael D. Westmoreland,
  ``Approximate Quantum Error Correction'',
  \emph{Quantum Inf. Proc.} {\bf 1}(1+2):5--12, 2002.

\bibitem{Aharonov-et-al} Dorit Aharonov, Michael Ben-Or, Russell Impagliazzo and Noam Nisan,
  ``Limitations of Noisy Reversible Computation'', 
  arXiv:quant-ph/9611028, 1996.

\bibitem{JungePalazuelos} Marius Junge and Carlos Palazuelos, 
  ``CB-norm estimates for maps between non-commutative $L_p$-spaces and quantum channel theory'',
  arXiv:1407.7684 [math.OA], 2014. 

\bibitem{M-HFW} Alexander M\"uller-Hermes, Daniel Stilck Fran\c{c}a and Michael M. Wolf, 
  ``Relative Entropy Convergence for Depolarizing Channels'',
  arXiv:1508.07021 [quant-ph], 2015. 
 
\bibitem{Markus-Gegenbeispiel} Markus Grassl,
  ``Entanglement-assisted quantum communication beating the quantum Singleton bound'', 
  talk at AQIS 2016, Taiwan.

\bibitem{Markus-Gegenbeispiel-Paper} Markus Grassl, 
  ``Entanglement-Assisted Quantum Communication Beating the Quantum Singleton Bound'',
    {\em Phys. Rev. A} {\bf 103}(2):020601, 2021.

\bibitem{QuShannonRT} Igor Devetak, Aram W. Harrow and Andreas Winter,
  ``A Resource Framework for Quantum Shannon Theory'',
  \emph{IEEE Trans. Inf. Theory} {\bf 54}(10):4587--4618, 2008. 
  
\bibitem{NadkarniGarani-2021}  
  Priya J. Nadkarni and Shayan Srinivasa Garani,
  ``Non-binary Entanglement-assisted Stabilizer Codes'',
  \emph{Quant. Inf. Proc.} {\bf 20}(8):256, 2021

\bibitem{LaiAshikhmin} Ching-Yi Lai and Alexei Ashikhmin, 
  ``Linear Programming Bounds for Entanglement-Assisted Quantum Error-Correcting 
  Codes by Split Weight Enumerators'',
  \emph{IEEE Trans. Inf. Theory} {\bf 64}(1):622--639, 2018. 

\bibitem{Allahmadi:2021}
  A. Allahmadi, A. AlKenani, R. Hijazi, N.Muthana, F. \"Ozbudak and P. Sol{\'e},
  ```New constructions of entanglement-assisted quantum codes'',
  \emph{Cryptogr. Commun.}, 2021. 
 
\bibitem{WildeBrun-2008} Mark M. Wilde and Todd. A. Brun,
  ``Optimal entanglement formulas for entanglement-assisted quantum coding'',
  \emph{Phys. Rev. A} {\bf 77}(6):064302, 2008.

\bibitem{Carlet-2018} Claude Carlet, Sihem Mesnager, Chunming Tang
  and Ruud Pellikaan, ``Linear Codes Over $\mathbb{F}_q$ Are Equivalent to LCD Codes for $q>3$'',
  \emph{IEEE Trans. Inf. Theory} {\bf 44}(4):3010--3017, 2018.

  \bibitem{LaiBrun2012} Ching-Yi Lai and Todd A. Brun,
    ``Entanglement-assisted quantum error-correcting codes with imperfect ebits'',
    \emph{Phys. Rev. A} {\bf 86}(3):032319, 2012.

\bibitem{Galindo2019} Carlos Galindo, Fernando Hernando, Ryutaroh Matsumoto and Diego Ruano,
    ``Entanglement-assisted quantum error-correcting codes over arbitrary finite fields'',    
    \emph{Quant. Inf. Proc.} {\bf 18}(4):116, 2019. 
    See \cite{Galindo2021} for corrections.

\bibitem{Galindo2021} Carlos Galindo, Fernando Hernando, Ryutaroh Matsumoto and Diego Ruano,
    ``Correction to: Entanglement-assisted quantum error-correcting codes over arbitrary finite fields'',    
    \emph{Quant. Inf. Proc.} {\bf 20}(6):216, 2021. 

\bibitem{Rains2} Eric M. Rains, ``Quantum weight enumerators'',
    \emph{IEEE Trans. Inf. Theory} {\bf 44}(4): 1388--1394, 1998.

\bibitem{HiguchiSudbery2000} Atsushi Higuchi and Anthony Sudbery,
  ``How entangled can two couples get?'',
  \emph{Phys. Lett. A} {\bf 273}(4):213--217, 2000.

\bibitem{HsiehWilde} Min-Hsiu Hsieh and Mark M. Wilde,
  ``Entanglement-Assisted Communication of Classical and Quantum Information'',
  \emph{IEEE Trans. Inf. Theory} {\bf 56}(9):4682--4704, 2001.

\bibitem{Fannes} Mark Fannes, 
  ``A continuity property of the entropy density for spin lattice systems'',
  \emph{Commun. Math. Phys.} {\bf 31}(4):291--294, 1973.

\bibitem{Audenaert} Koenraad M. R. Audenaert,
  ``A sharp continuity estimate for the von Neumann entropy'',
  \emph{J. Phys. A Math. Theor.} {\bf 40}(28):8127--8136, 2007.

\bibitem{Winter:Fannes} Andreas Winter, 
  ``Tight Uniform Continuity Bounds for Quantum Entropies: 
  Conditional Entropy, Relative Entropy Distance and Energy Constraints'',
  \emph{Commun. Math. Phys.} {\bf 347}(1):291--313, 2016. 

   
\end{thebibliography}
\end{document}